\documentclass[
 reprint,
 superscriptaddress,
 amsmath,amssymb,
 aps,
 prl,
 longbibliography,
 onecolumn,
 showpacs
]{revtex4-1}
\usepackage{graphicx}
\usepackage{dcolumn}
\usepackage{bm}
\usepackage[colorlinks,linkcolor=blue,citecolor=blue]{hyperref}
\begin{document}
\title{Optimization and robustness of topological corner state in second-order topological photonic crystal}
\author{Xin Xie}
\author{Jianchen Dang}
\author{Sai Yan}
\affiliation{Beijing National Laboratory for Condensed Matter Physics, Institute of Physics, Chinese Academy of Sciences, Beijing 100190, China}
\affiliation{CAS Center for Excellence in Topological Quantum Computation and School of Physical Sciences, University of Chinese Academy of Sciences, Beijing 100049, China}
\author{Weixuan Zhang}
\affiliation{Key Laboratory of advanced optoelectronic quantum architecture and measurements of Ministry of Education, School of Physics, Beijing Institute of Technology, 100081, Beijing, China}
\affiliation{Beijing Key Laboratory of Nanophotonics $\&$ Ultrafine Optoelectronic Systems, Micro-nano Center, School of Physics, Beijing Institute of Technology, 100081, Beijing, China}
\author{Huiming Hao}
\affiliation{State Key Laboratory of Superlattices and Microstructures, Institute of Semiconductors Chinese Academy of Sciences, Beijing 100083, China}
\author{Shan Xiao}
\author{Shushu Shi}
\author{Zhanchun Zuo}
\affiliation{Beijing National Laboratory for Condensed Matter Physics, Institute of Physics, Chinese Academy of Sciences, Beijing 100190, China}
\affiliation{CAS Center for Excellence in Topological Quantum Computation and School of Physical Sciences, University of Chinese Academy of Sciences, Beijing 100049, China}
\author{Haiqiao Ni}
\affiliation{State Key Laboratory of Superlattices and Microstructures, Institute of Semiconductors Chinese Academy of Sciences, Beijing 100083, China}
\author{Zhichuan Niu}
\affiliation{State Key Laboratory of Superlattices and Microstructures, Institute of Semiconductors Chinese Academy of Sciences, Beijing 100083, China}
\author{Xiangdong Zhang}
\affiliation{Key Laboratory of advanced optoelectronic quantum architecture and measurements of Ministry of Education, School of Physics, Beijing Institute of Technology, 100081, Beijing, China}
\affiliation{Beijing Key Laboratory of Nanophotonics $\&$ Ultrafine Optoelectronic Systems, Micro-nano Center, School of Physics, Beijing Institute of Technology, 100081, Beijing, China}
\author{Can Wang}
\affiliation{Beijing National Laboratory for Condensed Matter Physics, Institute of Physics, Chinese Academy of Sciences, Beijing 100190, China}
\affiliation{CAS Center for Excellence in Topological Quantum Computation and School of Physical Sciences, University of Chinese Academy of Sciences, Beijing 100049, China}
\affiliation{Songshan Lake Materials Laboratory, Dongguan, Guangdong 523808, China}
\author{Xiulai Xu}
\email{xlxu@iphy.ac.cn}
\affiliation{Beijing National Laboratory for Condensed Matter Physics, Institute of Physics, Chinese Academy of Sciences, Beijing 100190, China}
\affiliation{CAS Center for Excellence in Topological Quantum Computation and School of Physical Sciences, University of Chinese Academy of Sciences, Beijing 100049, China}
\affiliation{Songshan Lake Materials Laboratory, Dongguan, Guangdong 523808, China} 



\begin{abstract}
The second-order topological photonic crystal with 0D corner state provides a new way to investigate cavity quantum electrodynamics and develop topological nanophotonic devices with diverse functionalities. Here, we report on the optimization and robustness of topological corner state in the second-order topological photonic crystal both in theory and in experiment. The topological nanocavity is formed based on the 2D generalized Su-Schrieffer-Heeger model. The quality factor of corner state is optimized theoretically and experimentally by changing the gap between two photonic crystals or just modulating the position or size of the airholes surrounding the corner. The fabricated quality factors are further optimized by the surface passivation treatment which reduces surface absorption. A maximum quality factor of the fabricated devices is about 6000, which is the highest value ever reported for the active topological corner state. Furthermore, we demonstrate the robustness of corner state against strong disorders including the bulk defect, edge defect, and even corner defect. Our results lay a solid foundation for the further investigations and applications of the topological corner state, such as the investigation of strong coupling regime and the development of optical devices for topological nanophotonic circuitry.
\end{abstract}
\maketitle
\section{Introduction}
Topological photonics enables robust manipulation of light, including directional transport and localization with built-in protection against defects and disorders. In conventional photonic topological insulators, the boundary states with one-dimensional lower than the bulk are hosted, exhibiting robust transport against defects and sharp bends \cite{lu2014topological,ozawa2019topological,haldane2008possible,wu2015scheme,noh2018observation}. Such robustness has been demonstrated in various systems \cite{wang2009observation,hafezi2011robust,fang2012realizing,rechtsman2013photonic,hafezi2013imaging,
shalaev2019robust,he2019silicon,khanikaev2013photonic}, promoting novel approaches to one-way waveguide \cite{wang2009observation,hafezi2011robust,fang2012realizing,rechtsman2013photonic,hafezi2013imaging,shalaev2019robust,he2019silicon}, robust laser \cite{bahari2017nonreciprocal,bandres2018topological,harari2018topological,zeng2020electrically,yang2020spin,
zhong2020topological,st2017lasing,zhao2018topological} and chiral quantum optical interface \cite{barik2018topological,barik2020chiral,mehrabad2020chiral}. Recently, a new class of photonic topological insulators, higher-order topological insulators, has been proposed and realized, which supports lower-dimensional boundary states \cite{kim2020recent,schindler2018higher}. For instance, in 2D case, the second-order topological insulators possess 0D corner state. The second-order topological insulators can be formed by various mechanisms, such as the quantization of bulk quadrupole polarization \cite{benalcazar2017quantized,imhof2018topolectrical,serra2018observation,peterson2018quantized,mittal2019photonic,he2020quadrupole,
dutt2020higher}, the quantization of dipole polarization including the second-order topological insulators in a 2D kagome lattice \cite{ezawa2018higher,xue2019acoustic,noh2018topological,ni2019observation,kempkes2019robust,el2019corner,li2020higher} and 2D square lattice \cite{benalcazar2020bound,xie2019visualization,chen2019direct,ota2019photonic,cerjan2020observation} based on the generalized Su-Schrieffer-Heeger (SSH) model, and others \cite{liu2019second,luo2019higher,zhang2021experimental,jung2020nanopolaritonic}. In contrast to other mechanisms, the 2D generalized SSH model is easily implemented in all-dielectric photonic crystal (PhC), which has been demonstrated in various electromagnetic systems, ranging from the microwave \cite{xie2019visualization,chen2019direct} to the optical domain \cite{ota2019photonic}. Based on 0D corner state in the second-order topological PhC, a topological nanocavity can be formed. Different from the previous PhC cavities \cite{akahane2003high,song2005ultra,shirane2007mode}, the localization of corner state results from the global property of bulk band, i.e., topological bulk polarization, not the local change in properties introduced by the defect. Therefore, the localized property of corner state is topologically protected, exhibiting robustness to disorder and defects \cite{li2020higher,proctor2020robustness,benalcazar2020bound,coutant2020robustness}.

By coupling to solid-state light emitters, the topology-driven localization of optical states in active photonic structures open avenues for various applications, including cavity quantum electrodynamics (CQED) \cite{xie2020cavity} and topological nanophotonic devices such as topological nanolaser \cite{zhang2020low,han2020lasing,kim2020multipolar}. The performances of these applications are highly related to the quality factor (Q) of the corner state. However, in the topological nanocavity without optimization, the calculated Q of corner state is only in the order of 10$^3$ \cite{han2020lasing} and a maximum experimental Q (Q$_{exp}$) is about 2500 \cite{ota2019photonic}, limiting further applications. Therefore, the optimization of Q is essential for further investigation of corner state, which still lacks except for the method reported in our previous works by changing the gap between two PhCs \cite{xie2020cavity,zhang2020low}. Additionally, although the robustness of the corner state against disorders in the topological nanocavity has been discussed in previous works \cite{zhang2020low,han2020lasing,shi2021coupled}, they are mainly the theoretical discussions, which just focus on one kind of specific defect or perturbation. A systematic study on the robustness with the experimental demonstration in particular has been rarely undertaken.

Here, we systematically investigate the optimization and robustness of the topological corner state in the second-order PhC. Based on the 2D generalized SSH model, the topological PhC nanocavity is designed, supporting 0D corner state. The Q of the corner state is optimized by three methods, including tuning the gap between two PhCs, downsizing the nearest-neighbor airholes around corner and shifting the nearest-neighbor airholes away from corner. The topological cavities with different modulation parameters are fabricated by using a GaAs slab embedded with quantum dots (QDs) served as light sources. The optical properties are characterized by the photoluminescence (PL) spectroscopy. To improve the Q$_{exp}$, the surface passivation treatment is performed to reduce the surface absorption. Meanwhile, we theoretically and experimentally demonstrate the robustness of corner state against strong defects including the bulk defect, edge defect and corner defect, which are introduced by removing airhole in the bulk, edge and corner, respectively. Our results demonstrate the potential and strength of the topological corner state, enabling the further investigation of CQED \cite{Qian2018,Qian2019} and development of topological nanophotonic devices with diverse functionalities.

The remainder of this paper is organized as follows. In Sec. \ref{s1}, we present the design of topological cavity and calculation results of the corner state in the modulated cavities. In Sec. \ref{s2}, the fabrication process and measurement setup are described. In Sec. \ref{s31}, we present our measurement results of the cavity modes in the modulated cavities. The experimental results measured at room temperature (RT) and at low temperature (LT) are compared. In Sec. \ref{s32}, the properties of cavities before and after surface passivation treatment are demonstrated. In Sec. \ref{s4}, the robustness against bulk defect, edge defect and corner defect are discussed. Finally, a brief conclusion is provided.

\section{Design and optimization of topological nanocavity}\label{s1}
\subsection{Design of topological nanocavity}\label{s11}

\begin{figure}[t]
\centering
\includegraphics[scale=0.5]{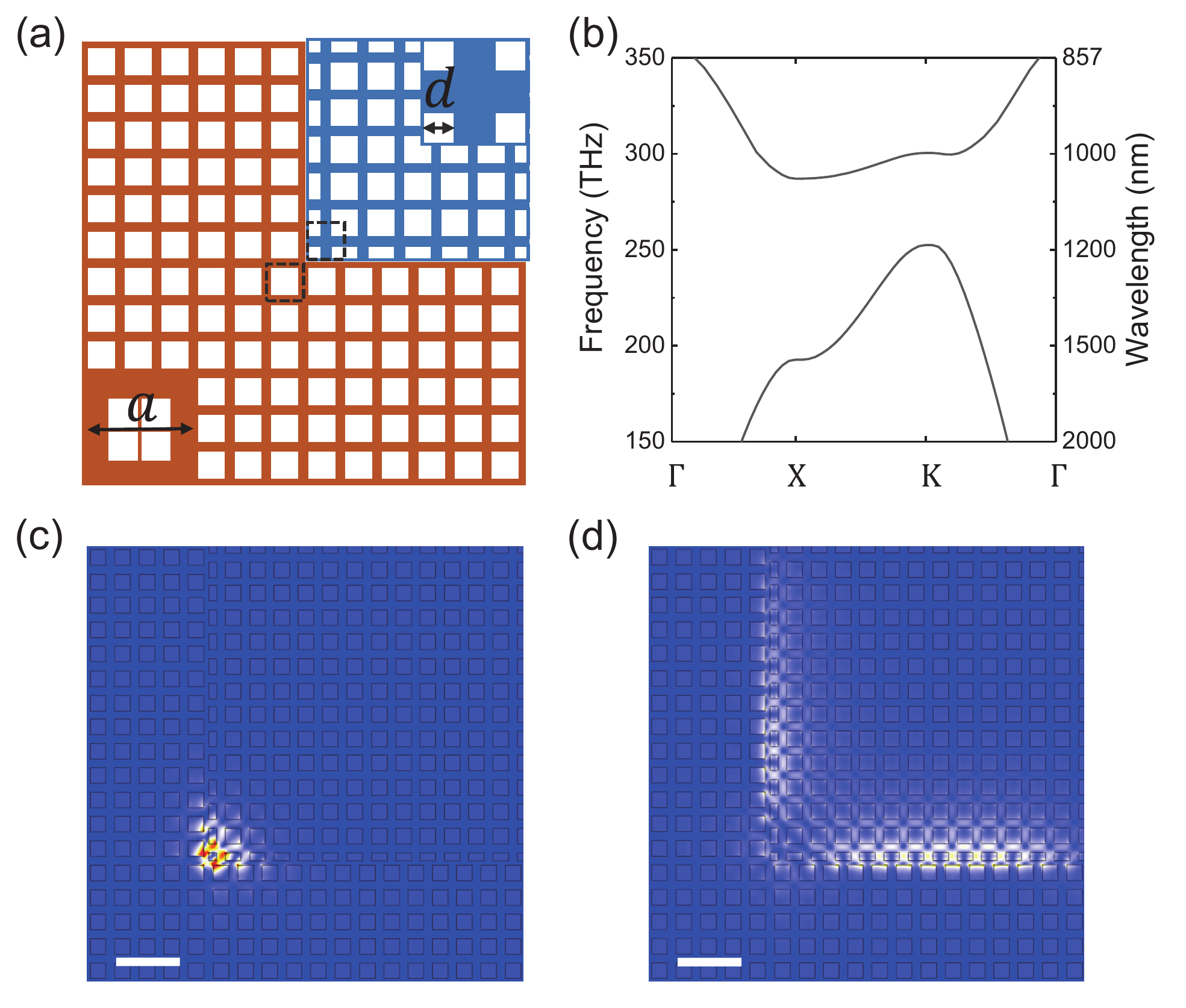}
\caption{(a) Schematic of 2D topological PhC cavity. The topological nanocavity consists of two topologically distinct PhCs in a square shape, PhC1 (colored in blue) and PhC2 (colored in orange). The white regions represent the airholes in PhC slab. Insets show the unit cells of the two PhCs with the same lattice constant $a$ and length of sub-airholes $d$. (b) Bandstructure of the PhC with $a=$ 380 nm, $d=$ 120 nm and n = 3.43, where n is the refractive index of the PhC's host material. (c) Electric field profile of topological corner state with a wavelength of 1131.12 nm and Q of approximately 3700. (d) Electric field profile of one of the topological edge states with wavelength of 1163.73 nm. The scale bars in (c) and (d) are 1 $\mu$m. The simulations were performed by 3D FDTD.  }
\label{f1}
\end{figure}

\begin{figure}[t]
\centering
\includegraphics[scale=0.5]{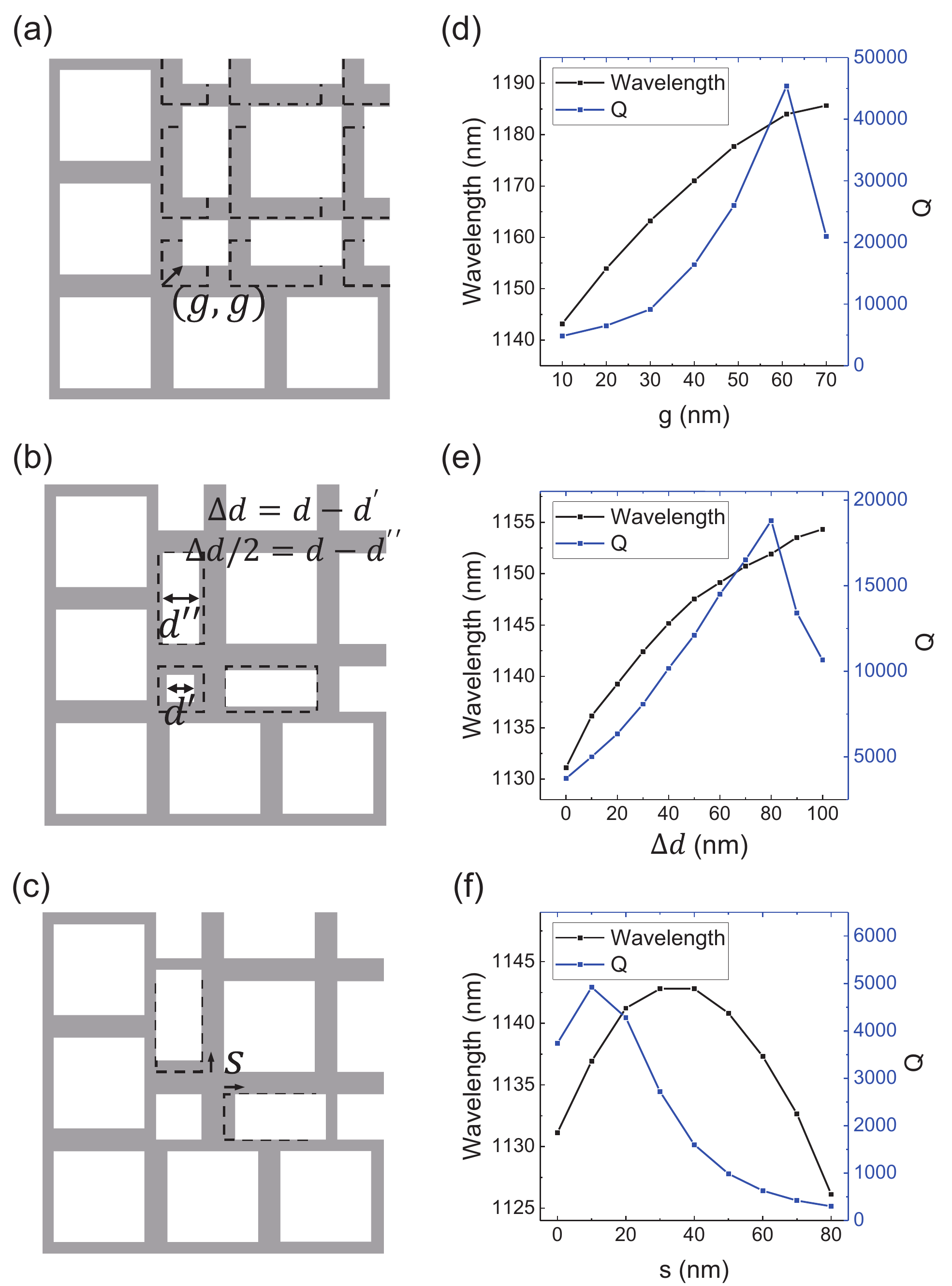}
\caption{(a)-(c) Schematics of modified topological cavity structures by (a) increasing the gap $g$, (b) downsizing the nearest-neighbour airholes at the corner and (c) shifting the nearest-neighbour edge airholes away from the corner. The dashed lines represent the unmodified case. (d)-(f) Calculated Q (blue lines) and resonance wavelength (black lines) of corner state as a function of modulation parameters, including $g$, $\Delta d$ and $s$ shown in (a)-(c), respectively. The simulations were performed by 3D FDTD with geometrical parameters of $a=$ 380 nm, $d=$ 120 nm and $t=$ 150 nm, where $t$ is the thickness of PhC slab.  }
\label{f2}
\end{figure}

Based on the 2D generalized SSH model, two PhCs are constructed. The insets of Fig. \ref{f1}(a) show the unit cells of the two PhCs, PhC1 (blue) and PhC2 (orange). They have the same lattice constant $a$ and share the common bandstructure, as shown in Fig. \ref{f1}(b). The bandstucture is calculated by the finite difference time domain (FDTD) method with $a=$ 380 nm, $d=$ 120 nm and n = 3.43. While, the two PhCs possess different topologies, which are characterized by 2D Zak phase, an integration of berry connection within the first Brillouin zone \cite{zak1989berry,liu2017novel}.  For the PhC1 and PhC2, the 2D Zak phases are ($\pi$, $\pi$) and (0, 0) respectively, corresponding to the second-order topological PhC and trivial PhC. At the interface between the two PhCs, a 90-deg corner and 1D edges are formed, as shown in Fig. \ref{f1}(a). According to bulk-edge-corner correspondence, 0D corner state and 1D gapped edge states are deterministically generated at the corner and edges, respectively. Figure \ref{f1}(c) and \ref{f1}(d) show the electric field distribution of the 0D corner state and one edge state calculated by the FDTD. In this case, the calculated Q of the corner state is about 3700, similar to that in previous works \cite{zhang2020low,han2020lasing}. Given the fabrication imperfection and large optical loss of bulk materials with emitters, the achievable Q for active PhC cavity is very low. Therefore, optimization of corner state is highly desired for further investigations and applications in CQED and topological nanophotonic devices.

\subsection{Optimization of corner state}\label{s12}
To optimize Q in general, the spatial variation of electric field distribution should be gentle, so that the Fourier transformation does not have components inside the leaky region \cite{akahane2003high}. Based on the idea, the Q of corner state is optimized by three methods to make the mode distribution gentler, including increasing the gap $g$ between the two PhCs, downsizing the nearest-neighbour airholes at the corner and shifting the nearest-neighbour edge airholes away from the corner. Figure \ref{f2}(a)-(c) show the schematics of the modified structures, in which the dashed lines represent the unmodified case. Figure \ref{f2}(d)-(f) show the corresponding calculated results of the corner state, including the resonance wavelength (black lines) and Q (blue lines), as a function of modulation parameters for these modified structures in Fig. \ref{f2}(a)-(c).

As the gap $g$ between the two PhCs is increased (Fig. \ref{f2}(a)), the corner state shows an apparent redshift and the Q first increases and then decreases, as shown in Fig. \ref{f2}(d). When g is equal to 60 nm, the Q reaches a maximum of about 46000, which is more than ten times larger than that in the unmodified cavity. Meanwhile, the corner state exhibits a large redshift of about 50 nm, and it is close to the band edge with large $g$. The result is similar to that in our previous works \cite{xie2020cavity,zhang2020low}, which is calculated by finite element method. Figure \ref{f2}(e) shows the calculated Q and wavelength of corner state as a function of $\Delta d$ for the modified structure in Fig. \ref{f2}(b), in which the length of the smallest square airhole is reduced by $\Delta d=d-d'$ and the length of the short side of the two nearest-neighbour edge rectangular  airholes are reduced by $\Delta d/2=d-d''$. As $\Delta d$ increases, the resonance wavelength increases, and the Q first increases and then decreases. The Q can be optimized by five times with a maximum about 19000 when $\Delta d$ is equal to 80 nm. In the case, the redshift is about 21 nm.  Figure 2(f) shows the calculated results as a function of $s$ for the modified structure shown in Fig. 2(c), where the two nearest-neighbour edge airholes are shifted away from the corner by $s$.  As $s$ increases, the wavelength first increases and then decreases. While, the Q first increases achieveing a maximum of only 5000 when $s=$10 nm, and then goes down very quickly. In contrast to the former two methods, the method by changing $s$ only exhibits a small increment in Q. Additionally, the latter two methods with parameters $\Delta d$ and $s$ exhibit smaller changes in wavelength and Q compared to that with parameter $g$. It may result from smaller influences on the modulation of electric field distribution, since only the nearest-neighbor airholes are modified in the two cases. Therefore, in addition to the method by changing the gap $g$, more delicate modulation around the corner can be used for optimization of Q.

\section{Fabrication and measurement}\label{s2}
Based on the calculation results, we fabricated these modified topological cavities with all the modulation parameters mentioned above into GaAs slabs. They contain a layer of InGaAs self-assembled QDs, which acts as broadband light sources to probe the optical properties of these topological cavities. The samples were grown by molecular-beam epitaxy on GaAs substrate. An 150-nm-thick GaAs slab with a layer of InGaAs QDs embedded in the center was grown on an 1-$\mu$m-thick Al$_{0.9}$Ga$_{0.1}$As sacrificial layer on the substrate. These modified structures were patterned into GaAs slab by electron beam lithography followed by inductively coupled plasma etching. The sacrificial layer below the structures was removed by wet etching with HF solution to form the air bridge.
\begin{figure}[b]
\centering
\includegraphics[scale=0.5]{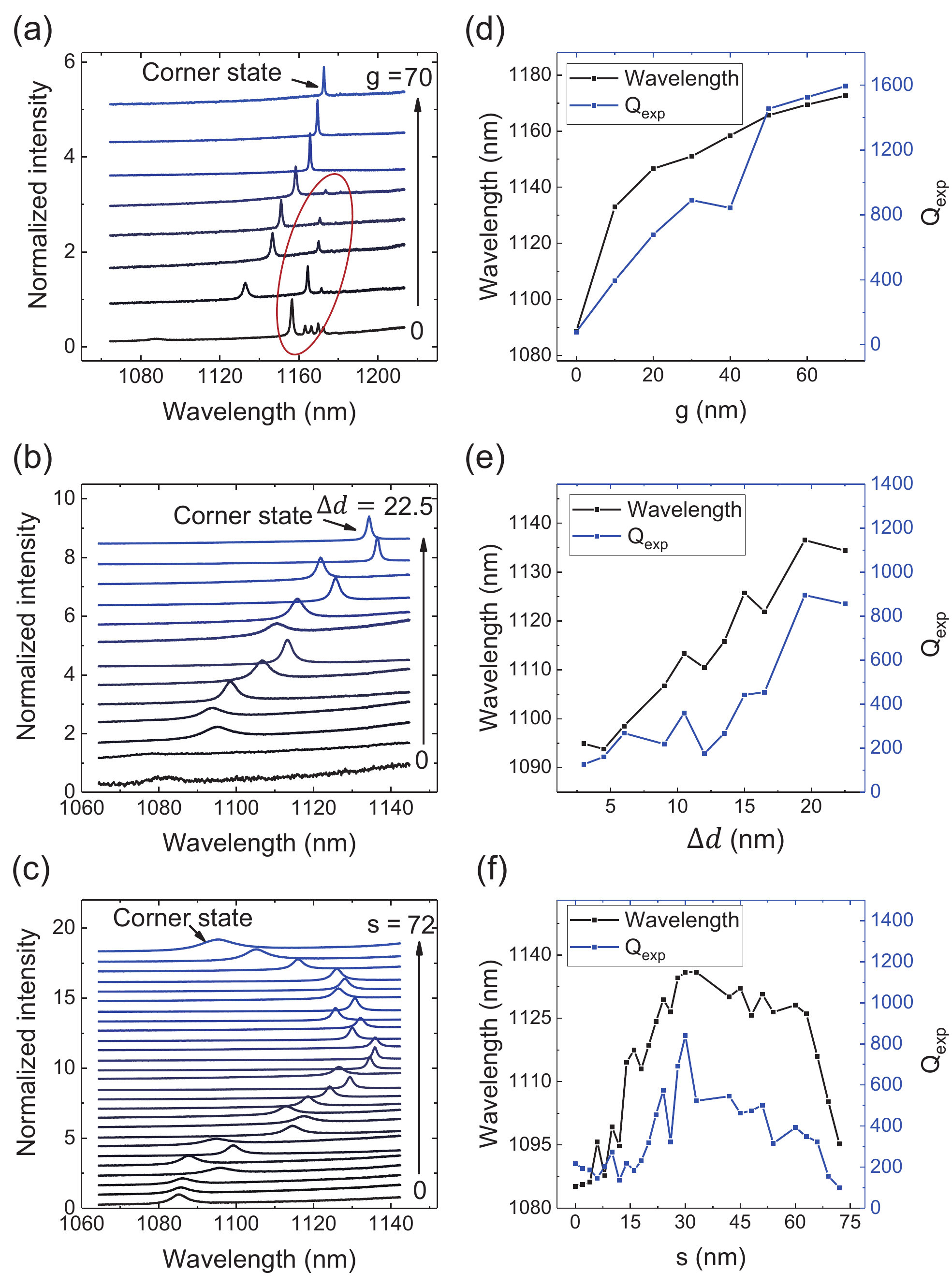}
\caption{(a)-(c) Normalized PL spectra in the modified cavities with different modulation parameters of (a) $g$, (b) $\Delta d$ and (c) $s$. The PL spectra were measured at RT. The PL peaks of corner state are indicated by the arrows. In (a), the PL peaks of edge states are also observed with small $g$, which have longer wavelength than that of the corner state, as indicated in the red region. The PL spectra are shifted for clarity. (d)-(f) Q$_{exp}$ (blue lines) and resonance wavelength (black lines) of corner state as a function of different modulation parameters, including (d) $g$, (e) $\Delta d$ and (f) $s$. The Q$_{exp}$  and resonance wavelength were extracted by Lorentz fitting with high-resolution spectra. The geometrical parameters for these cavities are $a=$ 380 nm, $d=$ 120 nm.  }
\label{f3}
\end{figure}

The confocal micro-PL measurements were performed at RT and LT (4.3 K) to characterize the optical properties of these cavities. The samples were cooled down to 4.3 K by a liquid helium flow cryostat. A 3D piezo-actuated translation stage was used for precise sample positioning. The samples were excited by a 532-nm continuous laser. An objective lens with a numerical aperture of 0.7 was used to focus the pump beam and collect the PL signal. The PL signal was then spectrally resolved by a grating spectrometer equipped with a liquid-nitrogen-cooled charge coupled device camera.

\section{Optical characterization}\label{s3}
\subsection{Experimental results of the modulated cavities}\label{s31}
\begin{figure}[b]
\centering
\includegraphics[scale=0.5]{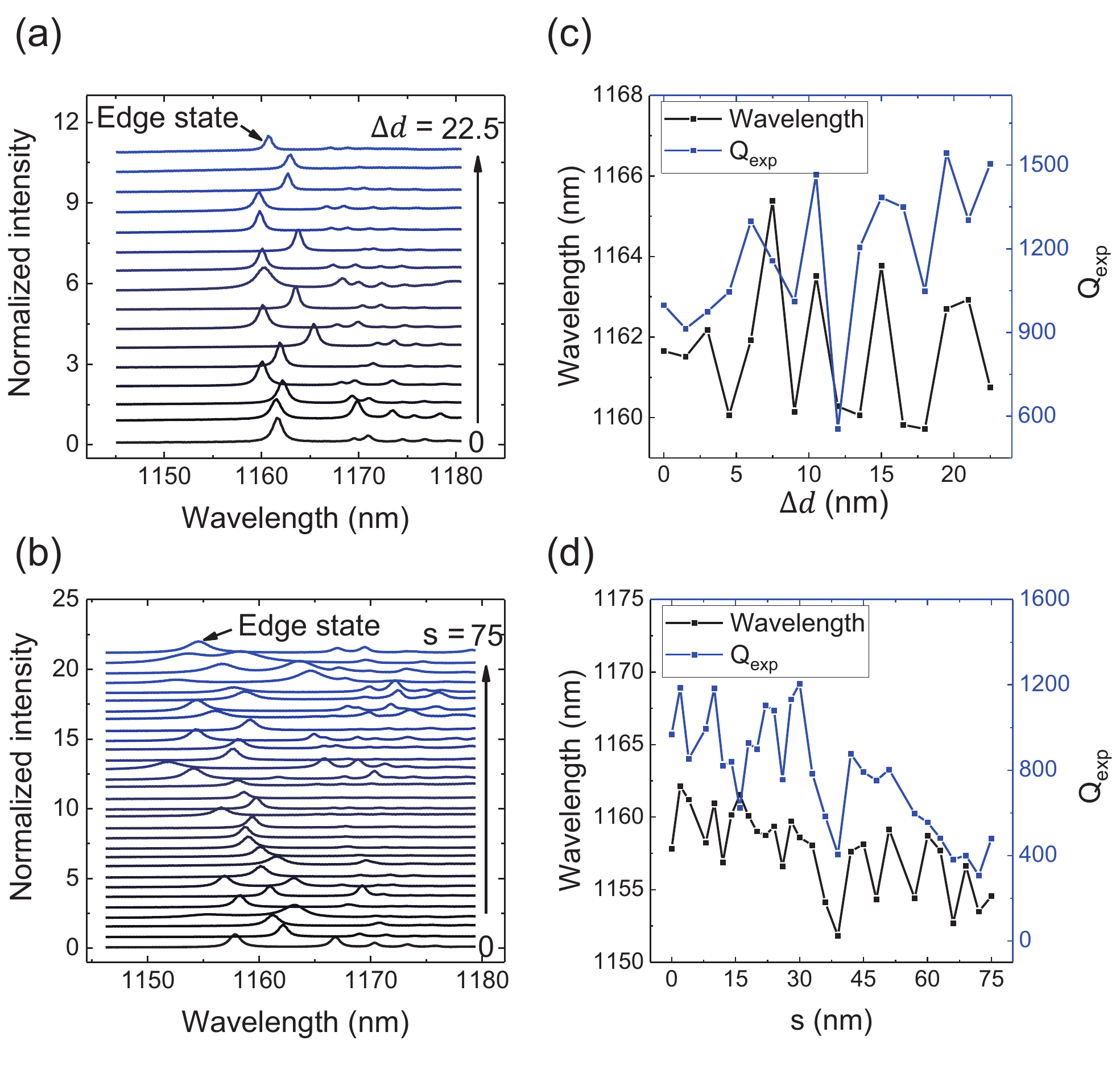}
\caption{(a)-(b) PL peaks of edge states in the modified cavities with different modulation parameters of (a) $\Delta d$ and (b) $s$. The PL spectra were measured at RT, and shifted for clarity. (c)-(d) Q$_{exp}$ (blue lines) and resonance wavelength (black lines) of one of edge states (indicated by the arrows in (a)-(b)) as a function of different modulation parameters of (c) $\Delta d$ and (d) $s$. The Q$_{exp}$ and resonance wavelength were extracted by Lorentz fitting with high-resolution spectra. }
\label{f4}
\end{figure}

Figure \ref{f3}(a)-(c) show the normalized PL spectra measured at RT in the modified cavities with different modulation parameters of $g$, $\Delta d$ and $s$. Their corresponding wavelength and Q$_{exp}$ of the corner state were extracted by Lorentz fitting to high-resolution spectra measured at RT, as shown in Fig. \ref{f3}(d)-(f). In Fig. \ref{f3}(a) and \ref{f3}(d), the corner state exhibits redshift accompanied with an apparent increase of Q$_{exp}$ as $g$ increases, agreeing well with the theoretical prediction shown in Fig. \ref{f2}(d). As $\Delta d$ increases, the overall trend of wavelength and Q$_{exp}$ is increasing as theory predicted (shown in Fig. \ref{f2}(e)). However, they have a relatively large fluctuation, as shown in Fig. \ref{f3}(b) and \ref{f3}(e). With $s$ increasing, the wavelength first increases and then decreases, in good agreement with the theoretical prediction (black line in Fig. \ref{f2}(f)) except for a small fluctuation. While, the Q$_{exp}$ show a large discrepancy with the theoretical results (blue line in Fig. \ref{f2}(f)). Meanwhile, in the three cases, the Q$_{exp}$ are much lower than the theoretical values due to the large optical absorption of host materials at RT and fabrication imperfection.

Additionally, PL peaks of edge states were also measured to investigate the influence of modulation parameters on the edge states. The red region in Fig. \ref{f3}(a) shows the PL peaks of edge states with different g, which have longer wavelength than the corner state. The edge states show apparent redshift with g increasing. When $g=$ 40 nm, the redshift is about 15 nm. Figure \ref{f4} shows the PL peaks of edge states with modulation parameters of $\Delta d$ and $s$. In contrast to the case with modulation parameter of $g$, the edge states are barely changed with modulation parameters of $\Delta d$ and $s$ since only the nearest-neighbour airholes around corner are modified. Meanwhile, the change of edge states in wavelength and  Q$_{exp}$ are much smaller than those of corner state, indicating edge states are more robust than corner state against perturbations around corner.

\begin{figure}
\centering
\includegraphics[scale=0.5]{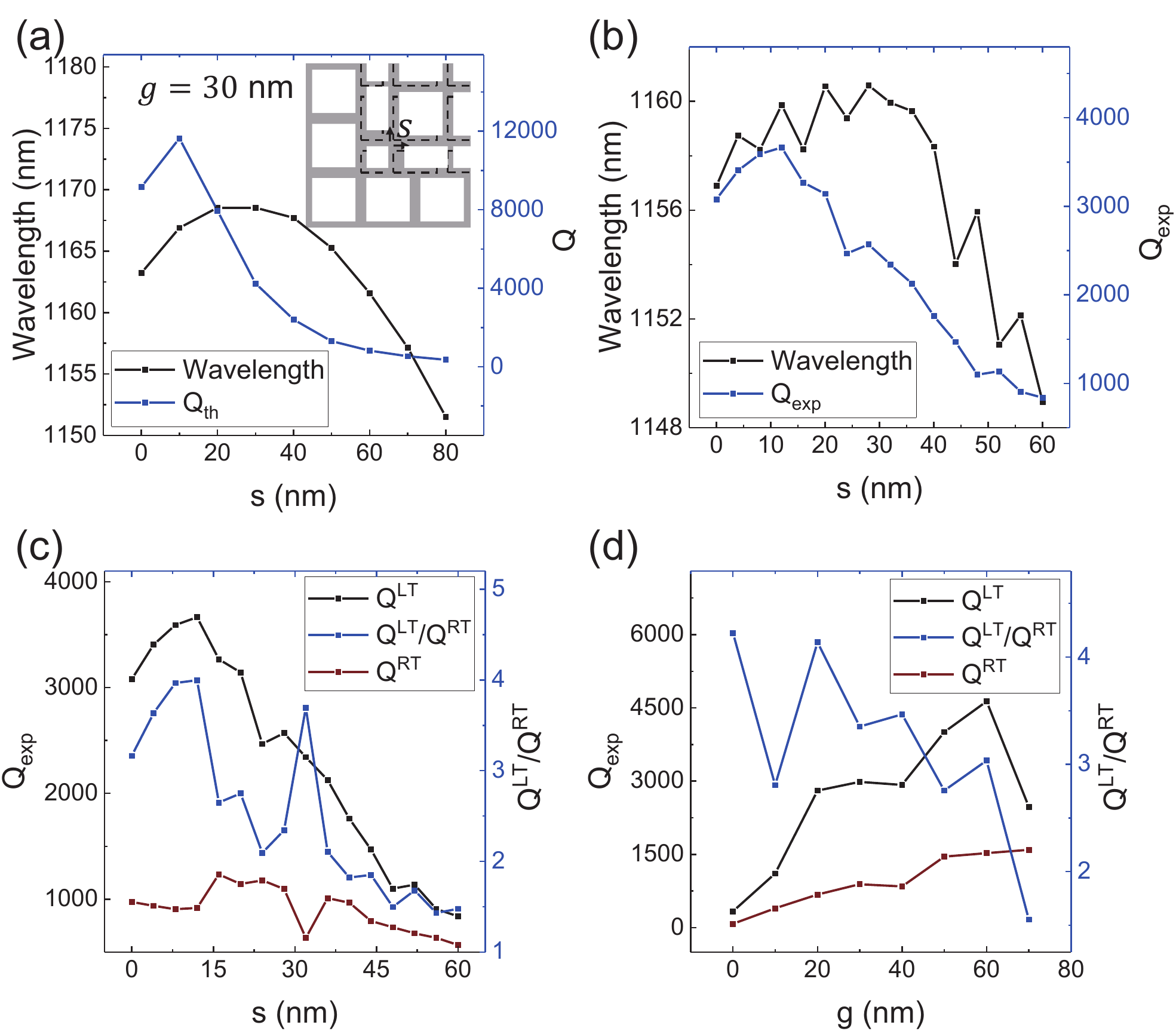}
\caption{(a) Calculated wavelength (black lines) and Q (blue lines) of corner state as a function of $s$ for the modified structure with $g=$ 30 nm. Inset shows the modified structure, in which the dashed lines represent the unmodified case. (b) Q$_{exp}$ (blue line) and wavelength (black line) of the corner state as a function of $s$, which were extracted by Lorentz fitting to high-resolution spectra at LT. (c)-(d) Comparison of Q$^{LT}$ (black lines) and Q$^{RT}$ (dark-red lines) as a function of (c) $s$ and (d) $g$, where Q$^{LT}$ and Q$^{RT}$ are the experimental results measured at LT and RT. The ratios between Q$^{LT}$ and Q$^{RT}$ are represented by the blue lines.  }
\label{f5}
\end{figure}

The discrepancy of wavelength and Q between theoretical and experimental results for corner state may result from the optical absorption and fabrication imperfection. On the one hand, the very strong optical absorption in active PhC cavities, especially at RT, leads to large optical loss, resulting in the reduction of Q$_{exp}$. On the other hand, the fabrication imperfection makes the fabricated cavities fluctuate in airhole sizes and positions, i.e., introducing structural disorders. The structural disorders lead to the fluctuation of wavelength. Note that the structural disorder becomes larger for smaller airholes. Moreover, light scattering from structural disorders may result in the reduction and fluctuation of Q$_{exp}$. 
Since the Q$_{exp}$ is influenced by the theoretical values, optical loss and fabrication imperfection, it can be described by:
\begin{equation}
\frac{1}{Q_{exp}}=\frac{1}{Q_{th}}+\frac{1}{Q_{ol}}+\frac{1}{Q_{fi}},
\label{e1}
\end{equation}
where Q$_{th}$ is the theoretical value, Q$_{ol}$ represents the influence of optical loss and Q$_{fi}$ represents the influence of fabrication imperfection. The influence of optical loss and fabrication imperfection will become greater with lower Q$_{th}$, leading to larger discrepancy between Q$_{exp}$ and Q$_{th}$. This explains the different degrees of discrepancy in the three methods.
To improve the Q$_{exp}$ and get a better coincidence with the theoretical results, a high Q$_{th}$ and low optical loss is necessary. Figure \ref{f5}(a) shows the calculated results as a function of modulation parameter $s$ in the modified cavity with $g=$ 30 nm. The change of the corner state with $s$ is similar to the case with $g=$ 0 nm, but with a higher Q. Meanwhile, in order to reduce the optical loss, we measured the PL spectra at 4.3 K (LT). The fitted results of Q$_{exp}$ (blue) and wavelength (black) are shown in Fig. \ref{f5}(b). The experimental results agree much better with the theoretical prediction, and the Q$_{exp}$ at LT is much higher than that at RT. Figure \ref{f5}(c) and \ref{f5}(d) show the comparisons of Q$^{LT}$ (black lines) and Q$^{RT}$ (dark-red lines) in the two cases with modulation parameters of $s$ and $g$. Apparent improvement of Q$_{exp}$ up to four times is observed due to the reduction of optical loss at LT, characterized by the ratio between Q$^{LT}$ and Q$^{RT}$ (blue lines in Fig. \ref{f5}(c) and \ref{f5}(d)). The different Q$^{LT}$/ Q$^{RT}$ in different cavities mainly result from the influence of fabrication imperfection and different values of Q$_{th}$.
\begin{figure}
\centering
\includegraphics[scale=0.5]{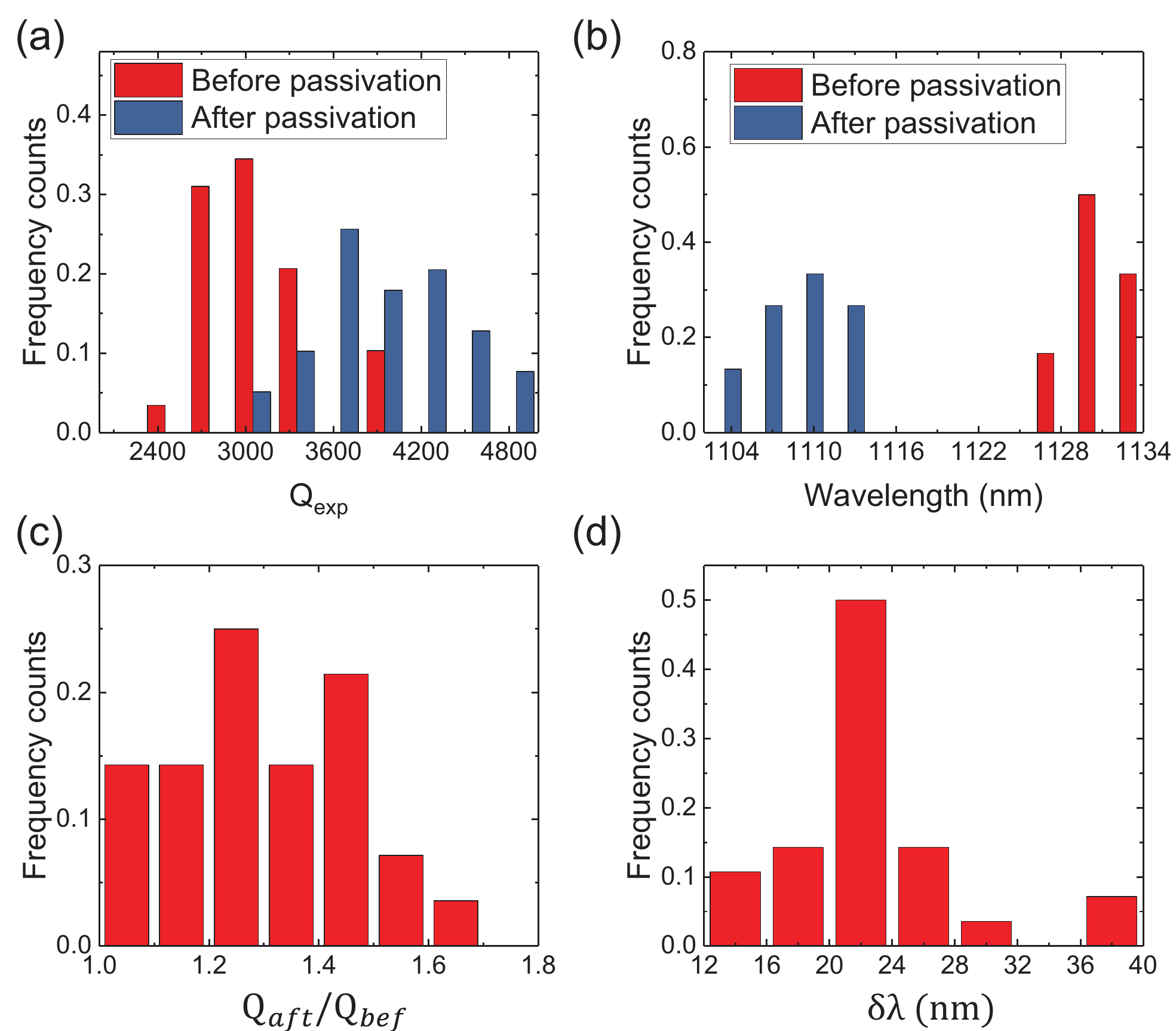}
\caption{Histograms of (a) Q$_{exp}$ and (b) wavelength for 42 cavities measured before (red) and after (blue) performing the surface passivation treatment, respectively. Histograms of (c) Q$_{aft}$/Q$_{bef}$ and (d) $\delta \lambda$ for the 42 cavities in (a)-(b), where $\delta \lambda=\lambda_{bef}-\lambda_{aft}$, Q$_{aft}$ and $\lambda_{aft}$ are the experimental results of passivated cavities, and Q$_{bef}$ and $\lambda_{bef}$ are the experimental results of the same cavities before passivation. }
\label{f6}
\end{figure}

\subsection{Surface passivation}\label{s32}

Besides decreasing the temperature, the optical loss can be further reduced by surface passivation treatment. The surface passivation treatment with Na$_2$S solution can suppress the light absorption at the surface of PhC’s host materials \cite{kuruma2020surface}. Here, we rinsed the samples into Na$_2$S solution for surface passivation treatment. First, we rinsed the samples into hydrogen peroxide solution for 30 s to oxidize the surface. Second, the samples were etched by citric acid for 1 min for surface cleaning. Finally, the samples were immersed into supersaturated Na$_2$S solution for 10 mins to passivate the surface. Before and after the surface passivation treatment, the PL spectra in the topological nanocavities were measured at LT.
\begin{figure}
\centering
\includegraphics[scale=0.45]{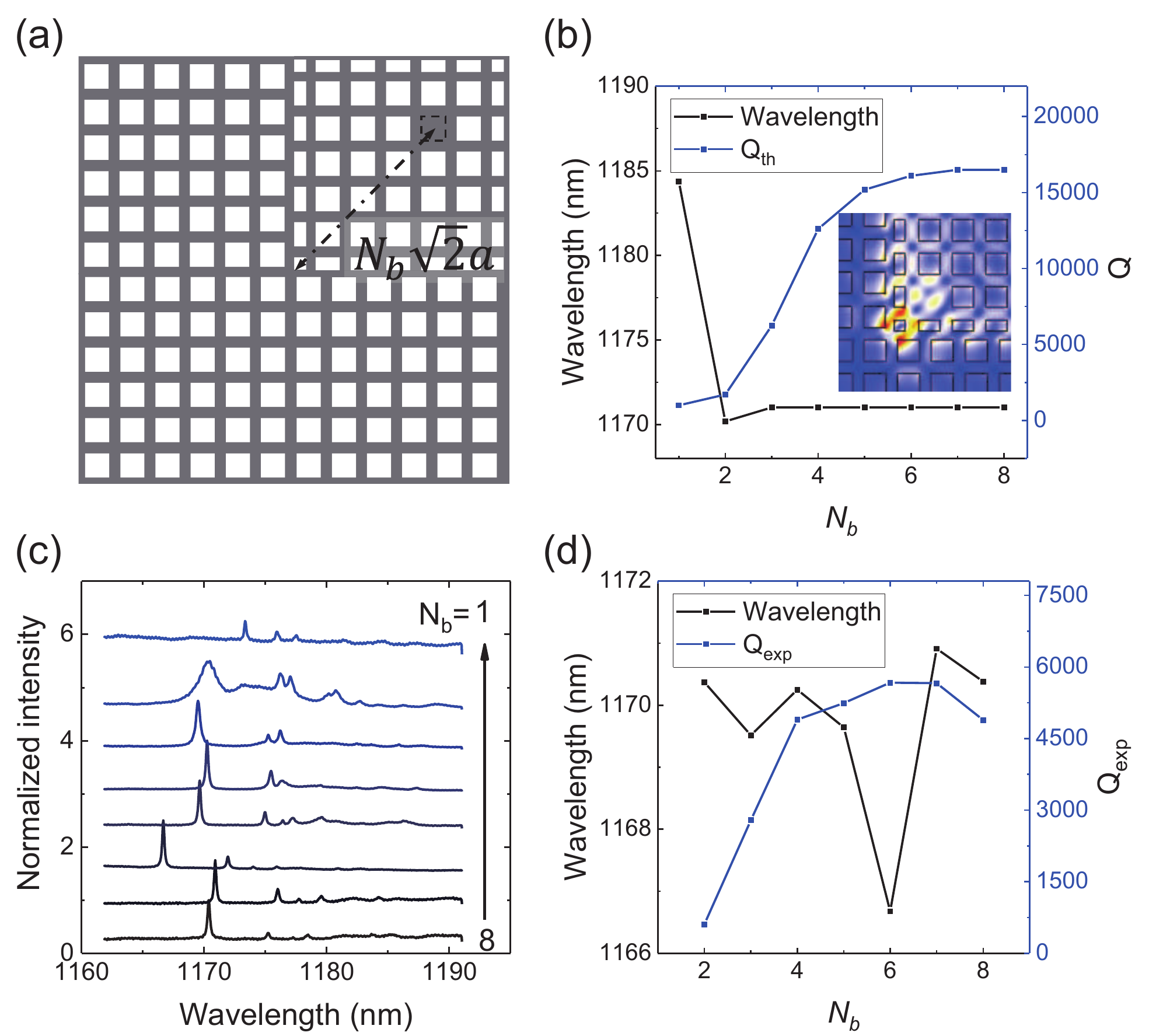}
\caption{(a) Schematic of topological cavity with bulk defect. The bulk defect is introduced by removing an airhole along the diagonal of PhC1. $N_b$ represents the distance of the bulk defect away from the corner. (b) Calculated Q (blue line) and wavelength (black line) of corner state as a function of $N_b$. Inset shows the electric field profile of corner state in the cavity with $N_b=$ 1. (c) Normalized PL spectra in cavities with bulk defect measured at LT. The numbers indicate $N_b$. The PL spectra are shifted for clarity. (d) Q$_{exp}$ (blue line) and wavelength (black line) as a function of $N_b$. The Q$_{exp}$ and wavelength were extracted from the spectra in (c). The geometrical parameters of these cavities for simulation and experiment are $a=$ 380 nm, $d=$ 120 nm, and $g=$ 40 nm.}
\label{f7}
\end{figure}
\begin{figure}
\centering
\includegraphics[scale=0.45]{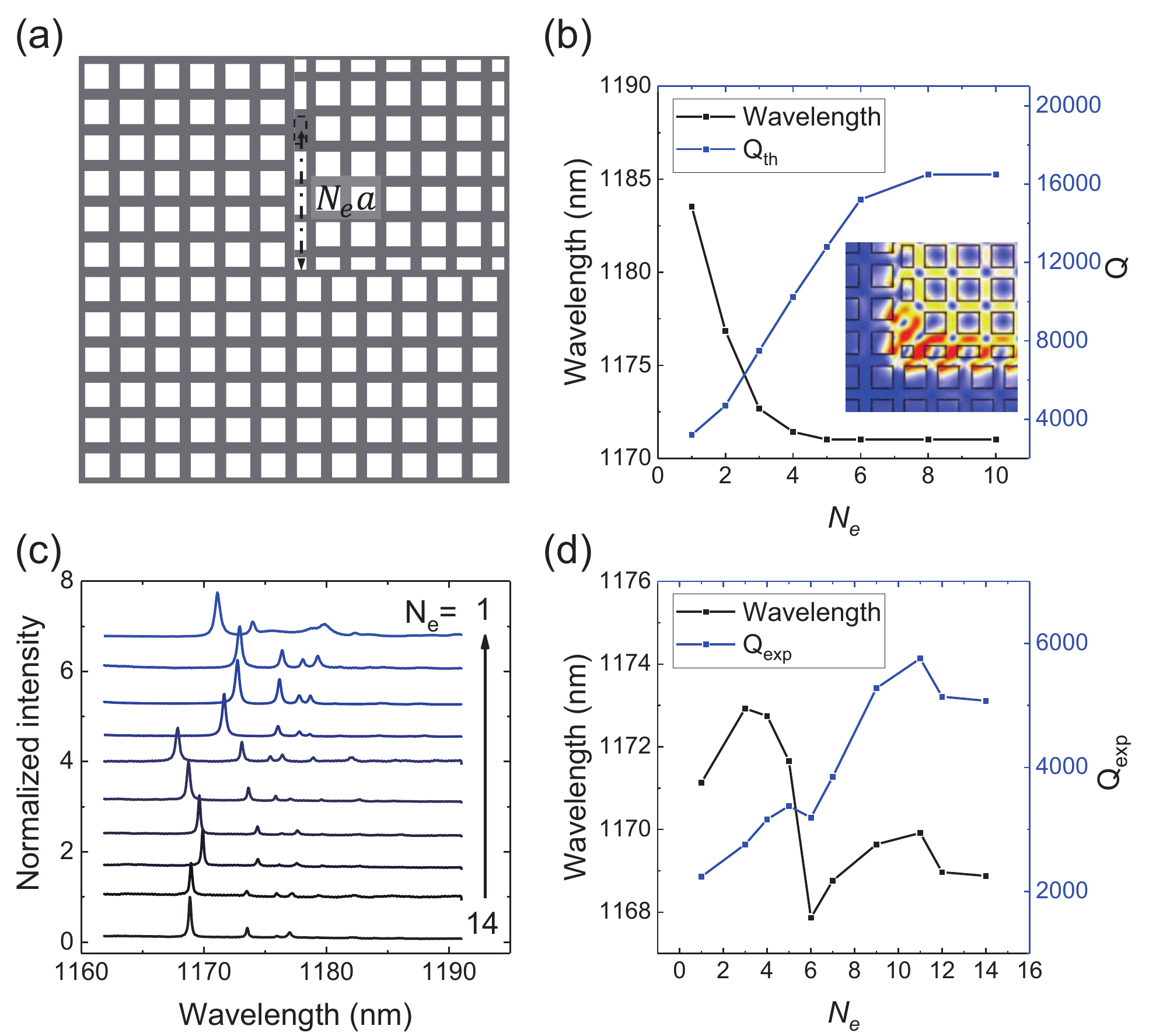}
\caption{(a) Schematic of topological cavity with edge defect, which is introduced by removing an airhole in y edge between PhC1 and PhC2. $N_e$ represents the distance of the edge defect from the corner. (b) Calculated Q (blue line) and wavelength (black line) of corner state as a function of $N_e$. Inset shows the electric field profile of corner state in the cavity with $N_e=$ 1. (c) Normalized PL spectra in cavities with edge defect measured at LT. The numbers indicate $N_e$. The PL spectra are shifted for clarity. (d) Q$_{exp}$ (blue line) and wavelength (black line) as a function of $N_e$. The geometrical parameters of these cavities for simulation and experiment are $a=$ 380 nm, $d=$ 120 nm, and $g=$ 40 nm.}
\label{f8}
\end{figure}

\begin{figure}[b]
\centering
\includegraphics[scale=0.5]{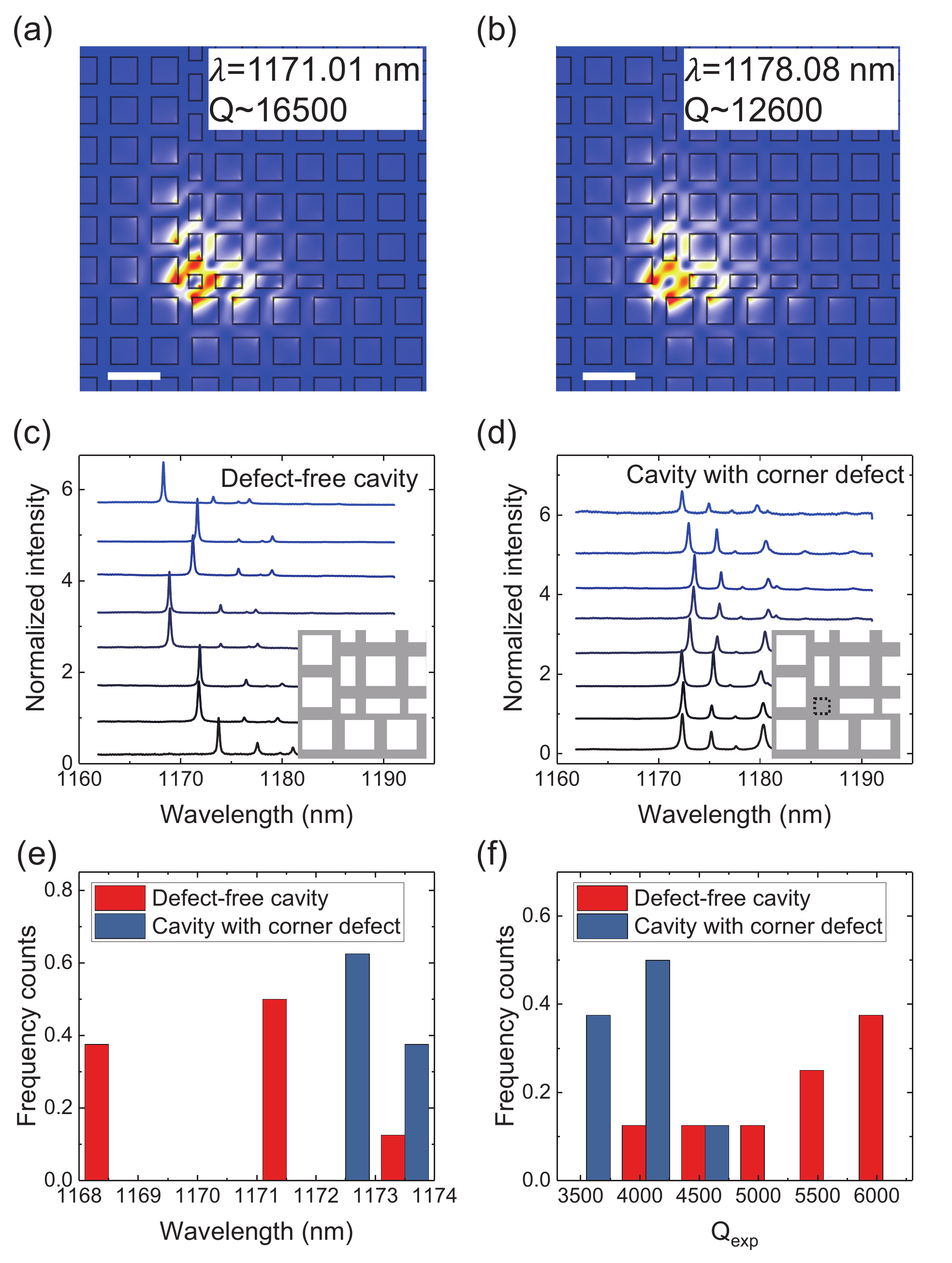}
\caption{(a)-(b) Electric field profile of corner state in (a) defect-free cavity and (b) cavity with corner defect. (c)-(d) Normalized PL spectra in (c) defect-free cavities and (d) cavities with corner defect measured at LT. Insets in (c)-(d) show the schematics of corresponding structures. The PL spectra are shifted for clarity. (e)-(f) Comparison of (e) wavelength and (f) Q$_{exp}$ between defect-free cavity (red) and cavity with corner defect (blue). The geometrical parameters of these cavities for simulation and experiment are $a=$ 380 nm, $d=$ 120 nm, and $g=$ 40 nm.}
\label{f9}
\end{figure}

Figure \ref{f6} shows the statistical results for the cavities measured before and after performing the surface passivation treatment. The results are obtained by the statistics of 42 cavities. Figure \ref{f6}(a) and \ref{f6}(b) show the frequency counts of Q$_{exp}$ and wavelength before (red) and after (blue) performing the surface passivation treatment, respectively. The passivated cavities exhibit higher Q$_{exp}$ and shorter wavelength. In order to quantify the improvement of Q$_{exp}$ and change of wavelength, we counted the ratio between Q$_{aft}$ and Q$_{bef}$ (Q$_{aft}$/Q$_{bef}$), and the difference of wavelength $\delta \lambda=\lambda_{bef}-\lambda_{aft}$, as shown in Fig. \ref{f6}(c) and \ref{f6}(d), respectively. Here, Q$_{aft}$ and $\lambda_{aft}$ are the experimental results of passivated cavities, Q$_{bef}$ and $\lambda_{bef}$ are the experimental results of the same cavities before passivation. The Q$_{aft}$/Q$_{bef}$ is in the range of about 1-1.7, demonstrating the improvement of Q$_{exp}$ by the surface passivation treatment, which results from the reduction of the surface optical absorption. The difference of wavelength $\delta \lambda$, is in the range of 12-40 nm, and mainly about 22 nm. The decrease in wavelength after passivation originates from the etching of GaAs slab and airholes by the citric acid and Na$_2$S solution.
\section{Robustness against defects}\label{s4}

The presence of topological corner state is topologically protected in a hierarchy of the bulk-edge and edge-corner correspondence as far as the 2D Zak phases of the bulk PhC are nontrivial in both directions. Therefore, the topological protection enables the robust existence of corner state even with harsh perturbations. Here, we numerically and experimentally investigate the effect of strong disorders on the wavelength and Q of corner state, including bulk defect, edge defect and corner defect.

We start with the investigation of robustness against bulk defect. The bulk defect is introduced by removing the airhole along the diagonal of the bulk of PhC1, as shown in Fig. \ref{f7}(a). The distance of the bulk defect from the corner is represented by $N_{b}$. Figure \ref{f7}(b) shows the calculated result of corner state as a function of $N_{b}$. As the bulk defect approaches the corner, the wavelength is barely changed except for $N_{b}=$ 1, and the Q shows an apparent decrease when $N_{b}<$ 5. The number $N_{b}$ required to maintain the same Q with the defect-free cavity is related to the localization length of the corner state. In the cavity with smaller localization length, the number can be reduced. For example, when $g=$ 10 nm, which have smaller mode volume \cite{xie2020cavity}, notable changes in Q occur when $N_{b}<$ 3. Although the Q is decreased by introducing very close bulk defect, the electric field of corner state is barely changed even with $N_{b}=$ 1, as shown in the inset in Fig. \ref{f7}(b). Figure \ref{f7}(c) shows the PL spectra for cavities with different $N_{b}$ measured at LT, and the corresponding fitted results of corner state are shown in Fig. \ref{f7}(d). As $N_{b}$ decreases, the wavelength is barely changed except for a fluctuation about 3 nm due to the fabrication imperfection, and the Q is decreased. The experiment result shows a similar tendency to the theoretical results. Therefore, the robustness of corner state against bulk defect is demonstrated by the almost constant wavelength and unchanged electric field distribution, even with very close bulk defect to the corner.

Then, we investigate the influence of edge defect introduced by removing arihole in the y edge, as shown in Fig. \ref{f8}(a). The distance of the edge defect from the corner is represented by $N_e$. Figure \ref{f8}(b) shows the calculated Q (blue line) and wavelength (black line) of corner state as a function of $N_e$. As the y-edge defect approaches the corner, the wavelength begins to increase when the edge defect is four periods away from the corner ($N_{e}<$ 5), and the Q decreases when $N_{e}<$ 7. Inset in Fig. \ref{f8}(b) shows the electric field of corner state when the nearest-neighbor edge defect is introduced ($N_{e}=$ 1). The distribution of corner state is still mainly located around the corner, similar to that in the defect-free cavity except the distribution is more dispersive. The experimental results are shown in Fig. \ref{f8}(c) and \ref{f8}(d). The PL spectra in Fig. \ref{f8}(c) were measured at LT, and the corresponding fitted results of corner state are shown in Fig. \ref{f8}(d). As $N_e$ decreases, a redshift in the wavelength and reduction of Q$_{exp}$ are observed, in agreement with the theoretical prediction. The little discrepancy of Q$_{exp}$ and wavelength between experiment and theory originates from the fabrication imperfection and optical absorption of the host material of PhCs. The corner state in the cavity with edge defect still exists and has the similar distribution with defect-free cavity even when the nearest-neighbor edge defect to the corner is introduced.

At last, we investigate the influence of corner defect by removing the smallest airhole at the corner, as shown in the inset of Fig. \ref{f9}(d). Figure \ref{f9}(a) and \ref{f9}(b) show the electric field of corner state in defect-free cavity and cavity with corner defect, respectively. They have very similar distribution despite the different values of wavelength and Q. The wavelength in cavity with corner defect is 1178.08 nm, longer than that in defect-free cavity (1171.01 nm). The Q in cavity with corner defect is about 12600, a little smaller than that in defect-free cavity (16500). Then we fabricated several defect-free cavities and cavities with corner defect possessing the same geometrical parameters, and characterized them by the PL spectra at LT, as shown in Fig. \ref{f9}(c) and \ref{f9}(d), respectively. Their corresponding wavelength and Q$_{exp}$ are counted, as shown in Fig. \ref{f9}(e) and \ref{f9}(f). In defect-free cavity, the Q$_{exp}$ are in the range of 4200-6200, and the wavelength are in the range of 1168-1174 nm. While, in the cavity with corner defect, the Q$_{exp}$ are in the range of 3400-4600, and the wavelength are in the range of 1172.3-1173.6 nm. In contrast to the defect-free cavity, the wavelength of cavity with corner defect is longer, and the Q is a little smaller, agreeing well with the theoretical prediction. Additionally, a reduced fluctuation of Q$_{exp}$ and wavelength in the cavities with corner defect is observed. It may be because the absence of the smallest airhole makes the influence of fabrication imperfection smaller. The structural disorder originating from fabrication imperfection becomes larger for smaller airhole, leading to a larger fluctuation in fabricated results. Therefore, in the defect-free cavity, the existence of the smallest airhole leads to a larger influence of fabrication imperfection than that in the case without the smallest airhole. The small difference between the defect-free cavity and cavity with corner defect is observed both theoretically and experimentally, demonstrating the robustness against corner defect.

\section{Conclusion}\label{s5}
In conclusion, we demonstrated the optimization and robustness of corner state both theoretically and experimentally. The Q is theoretically optimized from 10$^3$ to 10$^4$ by only modulating the nearest-neighbor airholes around corner or by changing the gap. Meanwhile, the Q$_{exp}$ is further optimized by surface passivation treatment due to the reduction of surface absorption. A maximum value of Q$_{exp}$ is about 6000, which is the highest value ever reported for the active topological nanocavity. Such high Q makes the strong coupling to single quantum emitters possible, paving the way for further applications. Moreover, the Q could be further optimized by more elaborately and flexibly tuning the position or size of not only the nearest-neighbor airholes, but also the second- and third-nearest-neighbor airholes, as far as the topological phase of the bulk PhC is not changed. We also demonstrate the robustness against bulk defect, edge defect and even corner defect both in theory and experiment, proving the strength of the topological cavity. The corner state still exists with similar field distribution even with strong defect which is very close to the corner, benefiting to the coupling to quantum emitters and applications for optical devices with built-in protection. Our results will accelerate the use of the topological cavity in various areas, such as the development of topological nanophotonic devices and investigation of the strong coupling regime, paving a way for far-ranging and widespread applications in topological nanophotonic circuitry.

\begin{acknowledgments}
This work was supported by the National Natural Science Foundation of China (Grants No.62025507, No. 11934019, No.11721404, No. 11874419 and No. 61775232), \textbf{the National key R$\&$D Program of China (Grant No. 2017YFA0303800 and No. 2018YFA0306101),} the Key R$\&$D Program of Guangdong Province (Grant No. 2018B030329001), the Strategic Priority Research Program (Grant No. XDB28000000) of the Chinese Academy of Sciences.
\end{acknowledgments}

\end{document}